# Towards Digital Twin-enabled DevOps for CPS providing Architecture-Based Service Adaptation & Verification at Runtime

Jürgen Dobaj[†]
Institute of Technical Informatics
Graz University of Technology
Graz, Austria
juergen.dobaj@tugraz.at

Andreas Riel
CNRS, G-SCOP Laboratory
Grenoble Alps University
Grenoble, France
andreas.riel@grenoble-inp.fr

Thomas Krug
Institute of Technical Informatics
Graz University of Technology
Graz, Austria
t.krug@tugraz.at

Matthias Seidl
Institute of Technical Informatics
Graz University of Technology
Graz, Austria
matthias.seidl@tugraz.at

Georg Macher
Institute of Technical Informatics
Graz University of Technology
Graz, Austria
georg.macher@tugraz.at

Markus Egretzberger
R&D Automation
Andritz Hydro GmbH
Vienna, Austria
markus.egretzberger@andritz.co

## ABSTRACT

Background: Industrial Product-Service Systems (IPSS) denote a service-oriented way of providing access to cyber-physical systems' (CPS) capabilities. The design of such systems bears high risk due to uncertainty in requirements related to service function and behavior, operation environments, and evolving customer needs. Such risks and uncertainties are well known in the IT sector, where DevOps principles ensure continuous system improvement through reliable and frequent delivery processes. A modular and service-oriented system architecture complements these processes to facilitate IT system adaptation and evolution.

Objective: This work proposes a method to use and extend the Digital Twins (DTs) of IPSS assets for enabling the continuous optimization of CPS service delivery and the latter's adaptation to changing needs and environments. This reduces uncertainty during design and operations by assuring IPSS integrity and availability, especially for design and service adaptations at CPS runtime.

Methodology: The method builds on transferring IT DevOps principles to DT-enabled CPS IPSS. The chosen design approach integrates, reuses, and aligns the DT processing and communication resources with DevOps requirements derived from literature.

Results: We use these requirements to propose a DT-enabled self-adaptive CPS model, which guides the realization of DT-enabled DevOps in CPS IPSS. We further propose detailed design models for operation-critical DTs that integrate CPS closed-loop control and architecture-based CPS adaptation. This integrated approach enables the implementation of A/B testing as a use case and central concept to enable CPS IPSS service adaptation and reconfiguration.

[†]Corresponding author

Conclusion: The self-adaptive CPS model and DT design concept have been validated in an evaluation environment for operation-critical CPS IPSS. The demonstrator achieved sub-millisecond cycle times during service A/B testing at runtime without causing CPS operation interferences and downtime.

## 1 Introduction

For several decades various industries have been dealing with the dynamic interplay of products and services known as Product-Service Systems (PSS) [1]. Market competition, higher profitability of services, and the need for more sustainable customer relations through increased and extended product lifecycle control are among today's top PSS drivers. These benefits led to PSS adoption between business partners, known as Industrial Product-Service Systems (IPSS) [2].

IPSS denote a service-oriented way of providing access to cyber-physical systems (CPS) capabilities. According to a very recent analysis by Brissaud et al. [3], remote asset maintenance and monitoring and remote asset reconfiguration are considered the most relevant CPS service capabilities at present and in the near future. Digital Twins (DTs) as virtual counterparts to these assets are accepted as an enabler for IPSS and the above use cases in particular.

Due to their vast business and sustainability potentials, IPSS have been expected to be widely adopted in multiple sectors. However, Brissaud et al. [3] found that IPSS are not as widespread as suggested in a former study [1]. In their analysis, the high risks imposed on IPSS service providers is among the main adoption challenges. These risks arise from planning and design uncertainties [4] related to unclear requirements of service behavior, operation environments, and evolving customer needs. This finding is supported by industry studies [3,5], concluding that the risk imposed by technical, user behavioral, and service provisioning uncertainties is evident and must be properly managed for CPS IPSS adoption.

As for the former challenge, we propose reducing provider risks and design uncertainties by adopting the DevOps lifecycle for CPS by enabling the continuous delivery, adaptation, and improvement of CPS (IPSS) services. To that end, we propose a self-adaptive CPS design model that maps the Information Technology (IT) concept of frequent design-build-deployment cycles to the Operation Technology (OT) domain of CPS IPSS. These design models are developed based on the systematic transfer of DevOps principles to the CPS OT domain, thereby integrating the DT as a shared knowledgebase to enable context-aware self-adaptive CPS characteristics. These characteristics ensure CPS IPSS service sustainability through design, technology, and function evolution. As the main contribution of this work, we propose a novel software service deployment strategy that enables the autonomous deployment and verification of service changes in the CPS OT domain at runtime.

The remainder of this work is organized as follows. Section 2 describes the background, discusses related work, and states the research questions and methodology. Section 3 derives design requirements for CPS DevOps. Section 4 explains the proposed design models. Section 5 evaluates and discusses the results. Finally, Section 6 concludes the presented work.

## 2 Background and Related Work

This section introduces DevOps and discusses the gaps in DevOps for CPS. DevOps is a collection of principles from computer science that signify the integration and collaboration of IT development with IT operations activities [6–8]. Such integration of software service development (Dev) and operations (Ops) enables the continuous building, testing, and deployment of software services upon each software design modification. Consequently, provided services continuously improve, evolve, and can be validated in "real" operation. While such a scenario is pertinent for IT environments, we must note that we are far from achieving the same for OT environments.

### 2.1 Value Creation through DevOps

We start our discussion on DevOps for CPS from a value-oriented perspective. To that end, we roughly divide the DevOps lifecycle into four value streams, as shown in Figure 1. (a) The creation of value in the (Design &) Dev space by developing and testing new and improved services. (b) The creation of value by the downstream delivery of these services to the operations environment for their operationalization. (c) The creation of value by using, maintaining, and monitoring the provided services in the Operation & Maintenance (O&M) space. (d) The creation of value by operations monitoring and the delivery of upstream feedback for data-driven strategy and design planning in the Design (& Dev) space.

The closed-loop implementation of these four value streams in CPS OT environments can only be achieved if the required CPS and OT capabilities (i.e., requirements) are well integrated across multiple DevOps lifecycle phases. In order to achieve an appropriate level of integration, we postulate that these capabilities must be taken into account during CPS design.

For example, particular key strengths of DevOps are the short (upstream and downstream) feedback cycles between Dev and Ops [8], which are enabled through rigorous operations monitoring and the frequent and reliable delivery and deployment of software services changes.

**Upstream Feedback.** To implement upstream feedback, the service monitoring capabilities need to be integrated into the service operation and monitoring phases. Second, response capabilities must be provided to ensure data and information transfer from Ops to Dev. Third, the received data needs to be integrated and managed for subsequent analysis and planning.

**Downstream Feedback.** To implement downstream delivery of releasable services, the service deployment capabilities need to be integrated into service operation to ensure CPS availability, integrity, and safety at all times. Therefore, downstream deployment needs to integrate autonomous monitoring and failure response capabilities into the CPS O&M space.

### 2.2 DevOps in the Light of DTs and CPS IPSS

Next, we want to take a brief look at the achievements of industry and research in CPS design, development, and operation, focusing on aspects that support the realization of DevOps value streams and feedback cycles.

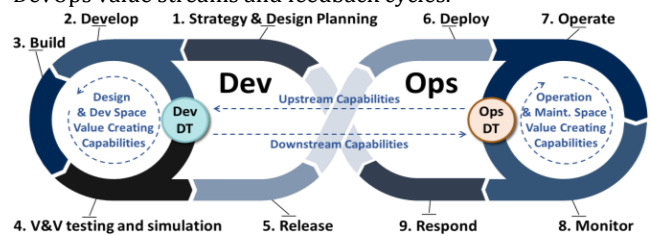

**Figure 1: DTs in the CPS DevOps Lifecycle**

**Digital Twin.** A DT consists of a physical entity, its virtual counterpart, and the data connection in between. Today, DTs are used to optimize physical entities in virtual space by experimenting with its virtual counterpart. The obtained results are then reflected back into physical space. Such a procedure is called twinning. The twinning fidelity describes the twinning accuracy determined by the twinned parameters and the twinning rate of these parameters [9].

**Design & Dev Space.** Traditionally, CPS were designed as special-purpose static silo applications, built and configured once for decades of operation with fixed pre-scheduled and manually executed maintenance intervals [10,11]. With the ever-increasing CPS complexity, due to the advancing digitalization and interconnection, and the need for value co-creation across industries [12], also CPS design has had to evolve. Self-adaptive systems engineering supports the management of this constant evolution by reducing the complexity and the uncertainty associated with change [13,14].

**Verification and Validation.** Today, DTs are used for shifting essential parts of the effort for predicting, evaluating, and preparing asset and CPS configurations virtually during design time rather than during expensive and availability-critical operation time [15]. For that purpose, model-based and data-driven simulation approaches are used [16–19]. When integrated into a framework, these approaches also enable virtual CPS engineering and commissioning [20] to reduce time, effort, and faults during physical commissioning.

**Connecting CPS Dev & Ops.** Despite great advances in (DT-based) simulation, a residual risk remains when deploying virtually validated services and functions to the real world. To further reduce CPS operation risks due to model and simulation uncertainty [20], verification approaches ensure that the virtual design space reflects physical reality accurately enough. Such verifications can be achieved by, e.g., linking the Ops DT (i.e., real world data) and the Dev DT (i.e., the virtual simulation space) to compare virtual with physical CPS behavior. To ensure interoperability and compatibility in case of evolving and diverging DTs, adaptive frameworks for DT data abstraction are proposed [21,22] to facilitate long-term upstream and downstream DT data exchange and data reuse.

**O&M Space.** In the operation space, DTs are used for remote operation and optimization of CPS assets and processes and predictive maintenance [23,24]. These DTs, e.g., minimize energy consumption and maximize asset lifetime and yield. Frequently such DTs rely on Model Predictive Control (MPC) facilitated through data-intensive machine-learning algorithms like Deep Neural Networks (DNN) [25]. Self-adaptive systems support CPS operations in, e.g., assuring dependability and resilience [26–29], and in adapting to changing environments and mission goals using, e.g., runtime models, goal-driven optimization techniques, and control theory [30–35], as well as to evolving software system needs by, e.g., relying on a modular design, architecture-based adaptation, and reuse [36–41].

**DevOps in the CPS Software Engineering Lifecycle.** Agile software practices, including DevOps, are increasingly adopted by regulated industries, where safety and standard compliance are cornerstones in software development for operation- and mission-critical CPS (used interchangeably in the following). These industries rely on rigorous engineering processes to deliver certifiable or certification-ready software. As concluded in [42], agile practices and CI/CD pipelines produce high-quality and certifiable software ready to be delivered more frequently to the end-user. However, such IT DevOps practices cannot be directly adopted due to stricter safety, compliance, and CPS architecture requirements. Leite et al. [8] made a similar observation when analyzing the DevOps concept and challenges in the IT sector. They note that adopting DevOps practices imposes significant challenges on system architecture, embedded systems, and IoT, including their design, management, and operation.

**The Vision of DevOps for CPS.** The increasing interest of regulated industry in DevOps practices resulted in proposals of concepts and roadmaps [43–49] pursuing the introduction of DevOps for CPS design and operations. Most of these works put the DT at the center of DevOps to serve as knowledgebase for organizing and automating DevOps processes in CPS. While all these and similar works we have found postulate that the DT can facilitate DevOps principles for CPS, we could not find any work that addresses CPS and DT service design for DevOps.

**Summing up,** significant progress has been made in all four value streams individually. In these value streams, the DT and self-adaptive systems engineering are becoming increasingly important to reduce uncertainty in the emergent and highly complex CPS landscape [50]. Particular strengths of today's DT-based CPS IPSS are their data-driven prediction capabilities and the DT's ability to manage and exchange data across lifecycle phases. However, these data exchange capabilities, especially today's downstream and O&M capabilities, do not support frequent and reliable software service deployment, a key aspect of DevOps in IT environment [8]. *To the best of the authors' knowledge, no work has addressed the research questions of adopting modern (IT-like) DevOps deployment strategies (DS) for critical CPS OT environments.*

### 2.3 Deployment Strategies and Architectures

In this section, we discuss the challenges of adopting IT DSs for the CPS OT environment. To that end, we describe state-of-the-art (SotA) IT DSs, whose characteristics are summarized in Table 1. We adapted this table from a presentation by the Cloud Native Foundation [51] and added the characteristics of SotA OT DSs. Figure 2 shows two conceptual architecture models to discuss the differences between IT and OT software service deployment and the IT and OT DevOps lifecycle processes.

**Architectural Aspects.** The most significant difference, when comparing the architectures in Figure 2, is that two networking technologies are used in the CPS architecture: (a) an IT network (NW) to support business and development processes and (b) an OT NW to support operations processes and to handle and respond to user requests (subsequently also denoted as user traffic (UT) and operations dataflow (ODF)). The air gap in Figure 2 indicates that these NWs are strictly separated, similar to the support processes implemented in these NWs. The traffic generated by these support processes (e.g., business, dev, and ops processes) is denoted as management traffic (MT) and management dataflow (MDF). The separation into traffic types is used throughout this work as an essential characteristic to describe DSs and the relations between users, supporting processes, service instances, and the information processing and communication (IPC) resources executing these services.

While there are many similarities between IT and OT NW technologies, there are also significant differences. Both NWs are designed for availability and high throughput rates. Since OT NWs are used to interact with the physical environment for critical application control, OT NW safety and integrity are essential for CPS operation. Further, the OT IPC resources and services must support real-time IPC in the sub-millisecond range to properly control "fast" physical processes. To meet these requirements, the OT IPC resources are organized in a strictly hierarchical manner (as indicated in Figure 2 by the IPC resource cluster). Resource-constrained embedded devices, for example, are located near IPC resource "users" to provide real-time control of physical assets and processes. Special purpose OT protocols and NWing technologies ensure safe and timely data transfer within and across OT NW hierarchy levels, called Automation Pyramid (AP) levels [52]. A detailed discussion of the AP and their impact on DevOps, DT, and CPS service design follows in Section 3. Next, we discuss different DSs.

**IT Deployment Strategies.** We start with the service DSs used in modern Cloud and IT IPC environments listed in Table 1. The simplest DSs is to recreate all service instances. In this case (a) the services running in the "current release" IPC cluster are stopped, and subsequently, (b) the new services are started in the same cluster. During this procedure, the system is not available to users. The deployment, usually fully automated, is managed by the individual DevOps team(s) responsible for the specific service. In an IT IPC environment, the MT and UT can be routed through a load balancer (LB) that inspects, filters, and forwards the different traffic types to their destinations. The LB's services are typically decoupled from the system service instances and operated in their individual IPC cluster.

The LB provides a standard interface to individual services, which enables the implementation of sophisticated DSs that, e.g., (a) ensure service availability during deployment (i.e., zero downtime), (b) enable testing of services and their comparison to current services in the real operation environments (i.e., real traffic testing), and (c) to perform the real traffic testing on specific users or user groups (i.e., targeted users). For example, shadow deployment allocates all the IPC resources to deploy all new service instances alongside the current release. 100% of incoming UT is routed to the current and new service releases, while only the responses of the current release are forwarded to the users. Hence, the shadow DS allows to compare releases causing zero downtime. The A/B testing DS, for example, allocates only IPC resources required for 30% of the overall user traffic. In this case, user requests are split and processed by the current (i.e., A) and new (i.e., B) service releases, as shown in Figure 2. The LB routes the responses from A and B to their respective users. Such a DS allows comparing user–reactions based on different service behavior. The LB's purpose is to manage and hide the complexity of these DSs.

**OT Deployment Strategies.** Implementing the above IT DSs and features in an OT environment is accompanied by several challenges. First, the strict real-time constraints and the interaction with distributed physical assets (represented by the users in Figure 2) prevent from using a "central" LB in OT environments. Instead, these constraints enforce a hierarchical structure that historically evolved to a NW of heterogeneous devices, protocols, and technologies. This structure and interaction with physical assets make IPC resources' dynamic allocation and management challenging [53]. As of Figure 2, it is feasible to realize only one dedicated IPC resource cluster to implement CPS functionality. Consequently, OT IPC resources are shared between (a) the Dev teams and processes, (b) the Ops teams and processes, and (c) the real-time and mission-critical CPS processes (i.e., the actual users). *Balancing all these "stakeholders" requirements and constraints makes designing and adopting DevOps and IT-like DSs for CPS challenging.*

| IPC ENV | Deployment Strategy | ZERO DOWN-TIME | REAL TRAFFIC TESTING | TARGETED USERS |
|---|---|---|---|---|
| Information Technology (IT) | RECREATE | ✗ | ✗ | ✗ |
| | RAMPED (aka Incremental, Rolling) | ✓ | ✗ | ✗ |
| | BLUE / GREEN (aka Red / Black) | ✓ | ✗ | ✗ |
| | CANARY | ✓ | ✓ | ✗ |
| | A/B TESTING | ✓ | ✓ | ✓ |
| | SHADOW (aka Mirrored, Dark) | ✓ | ✓ | ✗ |
| Operation Technology (OT) | MANUAL (LOCAL/REMOTE) UPDATE | ✗ | ✗ | ✓ |
| | (AUTOMOTIVE OTA) REMOTE UPDATE | ✗ | ✗ | ✓ |
| | **DT-enabled DevOps for CPS (this work)** | ✓ | ✓ | ✓ |

Table 1: Features of IT and OT Deployment Strategies

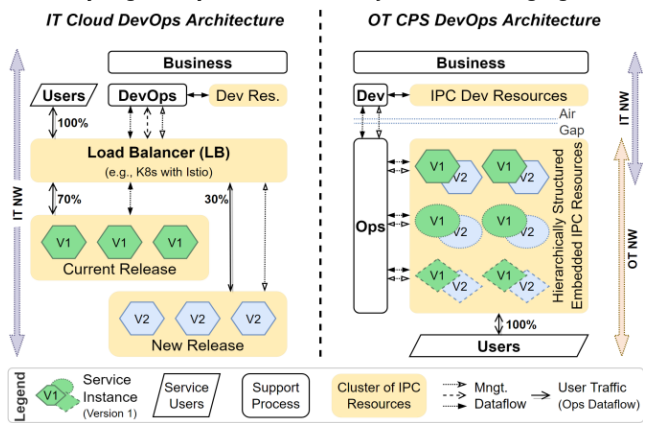

Figure 2: Conceptual IT and OT Deployment Models

Based on our experience in the field, we can say that most software updates are executed manually by dedicated and specially trained Ops teams (i.e., operation and commissioning engineers) that have local or remote access to the OT IPC cluster. This manual updating allows to easily apply changes to subsystems, i.e., targeted users as shown in Table 1. However, changes can typically only be applied if these subsystem or the entire CPS is taken offline (which results in heavy costs and high risks). Very recently, automated remote updates are becoming popular. Most notably, future vehicles are enforced by law to support software updates for security patching. The automotive industry addresses this via automated over-the-air (OTA) updates [54], the first step towards IT-like deployment.

**Summing up,** CPS are designed and developed as relatively static systems for a specific behavior and environment. Their maintenance is scheduled for fixed intervals, and significant IT-like hardware and software changes are usually not planned or supported only to a limited extent. Some sectors make their first steps towards more frequent software service updates. *However, to the best of the author's knowledge, no OT DS exists that supports modern deployment features such as real-traffic testing on targeted users at zero downtime.*

### 2.4 Research Questions and Methodology

The research objectives of this work are (a) to propose a self-adaptive service-oriented CPS model that uses the DT as a common knowledgebase throughout the DevOps lifecycle phases. Based on this model, we aim (b) to propose and evaluate the design of a novel A/B testing DS designed explicitly for OT environments to facilitate CPS DevOps by the features summarized in Table 1. From these objectives, we derive the following research questions (RQs), which contribute to the question of "*how CPS and its DTs shall be designed to realize DevOps for CPS OT environments*":

**RQ1** *What are the design requirements of self-adaptive CPS IPC services and their DT(s) to enable IT-like service deployment?*

**RQ2** *How shall DTs be structured to create a high-fidelity context for autonomous CPS adaption?*

**RQ3** *What does a DT-enabled self-adaptive CPS's conceptual model (aligned to [55] Fig. 1) look like?*

**RQ4** *Can the model covering RQ3 be instantiated to implement the proposed A/B testing for CPS?*

In the article presented here, we address *RQ1*, *RQ2*, and *RQ3* through the following methodology: We derive and analyze the design requirements and models covering system and device scale based on (a) previous work [53,56], (b) the integration of design patterns for self-adaptive systems and CPS engineering [57–60], and (c) the analysis and integration of concepts from several industrial reference architectures and frameworks [61–67]. We align the resulting requirements and models to the AP [52] underlying industrial CPS (see Figures 3 and 4). Subsequently, we address *RQ4* by proposing design models that facilitate the implementation of A/B testing for CPS. For validation, we implement and validate the A/B testing DS on an evaluation platform for distributed mission-critical CPS IPSS.

## 3 CPS and DT Design

In this section, we address the first three (design-related) RQs. To that end, we have compiled Figure 3, which shows multiple conceptual viewpoints on CPS aspects with a strong focus on the inherent hierarchical structure and organization of CPS.

### 3.1 Structure and Organization

**CPS IPSS Structure.** All views (a-f) that are shown in Figure 3 are vertically aligned to the AP levels. This alignment shall reflect that all views are strictly related to each other and that individual design aspects within a single view may impact the design and function of other views. In particular, view (a) maps the four DevOps value streams to (b) the AP levels. View (c) shows which IPC technologies are used on the individual AP levels, and (d) shows the typical physical layout of a CPS IPSS and its IPC clusters. View (e) maps the IT and OT deployment models as of Figure 2 onto the AP levels. Finally, view (f) shows the Dev DT and the Ops DT which are compositions of the individual DT instances distributed across the AP levels. The tiny boxes in view (d) indicate that these DT instances are further separated and distributed across the IPC NW resources.

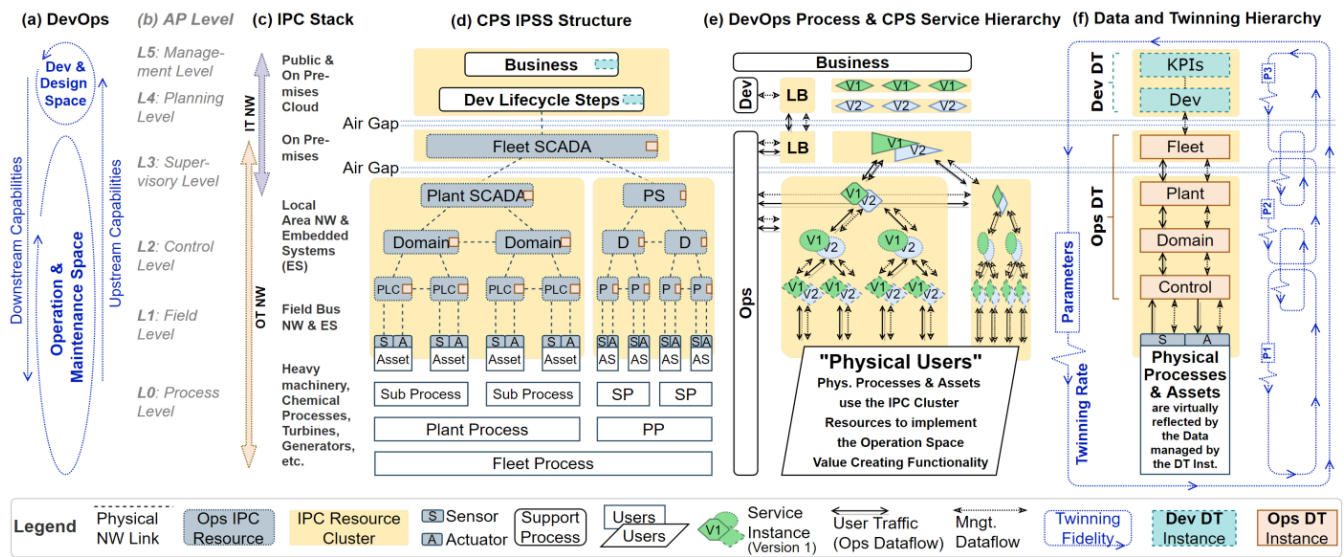

Figure 3: Views Showing Conceptual Models of the Hierarchical CPS Structure from a DevOps, Service, and DT Perspective

**IT NW.** AP level 5 is called business or management level. At this level, the DT provides key performance indicators (KPIs) of the overall CPS IPSS to drive business decisions. On level 4, the planning level, DTs are used for strategic system operations and design planning, development, simulation, and verification & validation. Due to their functional aspects levels 5 and 4 can be mapped to the Design & Dev space value stream. Modern IT solutions such as public and on-premises clouds are used for information processing and exchange at these levels.

**OT NW.** Level 3 denotes the supervisory level responsible for supervisory control and data acquisition (SCADA) of a plant (fleet). Level 2 is the control level, responsible for domain and asset control. In particular, the programmable logic controllers (PLCs) are connected via the field level technologies (i.e., sensor, actuators, fieldbus networks on level 1) to the physical assets and processes under control on level 0. The domain controllers are intermediate layers to increase CPS fault tolerance and ease the coordination of distributed assets and processes. The lower AP levels implement the mission-critical IPC and CPS functions. Hence, they are designed for availability, integrity, safety, and time-synchronous cyclic real-time IPC.

**CPS Functionality and Services.** The IPC and CPS functions are actually implemented by the distributed service instances shown in Figure 3(e), which are executed on the IPC resources (a) to coordinate and control the physical assets and processes (i.e., the "physical users" of the cluster), (b) to coordinate and implement the support processes, and (c) to manage IPC cluster resources and their usage. *The composition and cooperation of these IPC service instances within and across AP levels are essential for implementing the CPS functionality.*

**The DT as Enabler Technology.** As of Section 2, the DT serves as common knowledgebase in the Dev and Ops space. We propose this DT knowledgebase as enabler technology to ensure proper composition, cooperation, and management of the service instances and their deployment across AP levels. To that end, the envisioned DT must provide a context that reflects the physical assets, the physical processes, and the IPC resources to be monitored, maintained, (re)configured, and updated throughout the DevOps lifecycle. That DT context must be accurate enough at all times for (a) decision-makers to make the right decisions remotely, (b) for developers to deploy and validate those decisions in virtual (design) space (c) for developers and operators to deploy and validate the releasable design in the physical (operation) space (i.e., real traffic testing), and (d) to then close the loop by operationalizing the virtually and physically validated design in the physical space.

The Dev DT and Ops DT layouts (shown in Figure 3(f) and explained above) provide such an accurate DT-enabled context by facilitating individual twinning fidelities within and across AP levels. For example, the control DT instance is a high-fidelity real-time representation of physical asset and process states to implement proper closed-loop control. On the higher AP levels, the twinning fidelity of the same states can be reduced to, e.g., KPIs on level 5 describing the number of produced goods. This fidelity adaptation is achieved by data preprocessing and filtering between the AP levels, which is necessary to meet the different IPC requirements that change between AP levels. The adaptive twinning fidelity between AP levels also allows addressing different computational needs of twinning, (model-based) prediction, and simulation. Overall, the separation of the DT into a layered Dev DT and Ops DTs simplifies DT instance and communication design because only the requirements of specific AP levels have to be considered.

### 3.2  Design Requirements

**Paradigm Change.** Traditionally, users (learned to) adapt their behavior to the provided system functions. Nowadays, the system learns from the user and continuously adapts its service to user behavior, maximizing value for all stakeholders. This paradigm shift in system behavior is introduced with the IPSS concept in CPS, which must be accompanied by a paradigm shift in CPS design. Hence, modern CPS IPSS are no longer designed as static, siloed, special purpose systems. Instead, they need to become increasingly dynamic to adapt to changing needs, environments, and technologies. As of Sections 1 and 2, this requires CPS OT design to integrate agile processes and DSs.

**Abstraction & Interoperability.** From a technology point of view, one of the big challenges in OT design is that mutually largely incompatible networking concepts control different levels in the AP. Even between the same technologies, there might exist vendor-specific incompatibilities. Hence, a lot of manual integration and configuration work is required for design changes. The standardization and introduction of new fieldbus technologies and industrial Ethernet shall address this issue by using the IP suite as an enabler for data exchange across AP levels, making the traditional AP structure flatter and easier to handle [52]. However, according to Wollschläger et al. [11] this will not decrease the prevalent heterogeneity of communication protocols and technologies. Instead, heterogeneity will rather be set to increase in the future. Therefore, the *harmonization of the CPS IPSS services above the networking and communication layer is essential to ensure vertical and horizontal interoperability* (requirement R1).

**Autonomy.** The analysis presented in [56] states that the need for adaptation in CPS may arise due to technology advances, standardization, dynamic device integration unknown at the design time, and due to security flaws, which might result in changes of the communication infrastructure, protocols, or the software architecture itself. Hence, *CPS must also be capable of adapting to future changes in its system requirements and its technology stack (R2)*. As described in Section 2, such CPS adaptations are commonly performed manually. However, these adaptations must be automated using CI/CD pipelines [8] to effectively achieve CPS DevOps. At the same time, the *deployment to the operations space must be autonomous (i.e., no human-in-the-loop) to ensure CPS availability and integrity (R3)*.

**Service-Orientation & Modularity.** To establish autonomous adaptation across AP levels, the strict separation of management (e.g., IPC service and network configuration, monitoring, and arbitration) and operation (e.g., asset control and CPS monitoring) concerns becomes essential (a) to provide freedom from interference, (b) to cope with design and adaptation complexity, and (c) to ensure system evolution through modularity. Such a separation addresses technological evolution and interoperability and enables the independent development and evolution of operations and management functions. Therefore, *management and operation traffic (i.e., data flows) and management and operation services shall be strictly separated. Interactions between operation and management shall be minimal and implemented by well-controlled interfaces (R4)* [53].

**Context.** From an adaptation point of view, the hierarchical control pattern described in [57] is well suited to coordinate the above described decentralized, hierarchical, multi-step, and multi-technology CPS adaptations. DTs already available in CPS IPSS can be reused and extended to serve as knowledge repositories for contextual and coordinated adaptation. In particular, the Ops DT instances can provide the high-fidelity context-awareness to maintain CPS availability and integrity by: (a) ensuring that adaptations are only made in non-critical CPS states; and (b) canceling or delaying adaptations early if CPS performance is to be violated. So-called management DTs can run aside Ops DTs to maintain system integrity by creating self-awareness about IPC service, resource, and adaptation states required for coordinated adaptation across all AP levels.

### 3.3 DT-enabled Self-Adaptive CPS Model

This section introduces the conceptual model of a DT-enabled self-adaptive CPS (see *RQ3*) to establish a common ground for the subsequent development of detailed CPS service design models and adaptation strategies. Figure 4 shows our proposed adaptation model that is aligned to the conceptual self-adaptive system model described in [55] Fig. 1.

**DT-enabled Self-Adaptation**. The managing and managed subsystem boundaries in Figure 4 show the fundamental architecture of a self-adaptive system. The *managed system* (i.e., the entire CPS physical instance) is comprised of the SCADA and control services responsible for plant (fleet) control and physical system control. For that purpose, these services are executed on the OT IPC resource cluster on field, plant, and fleet levels. Sensors are used to replicate the physical system and process states into the (virtual) services for real-time closed-loop control. The actuators replicate the results back to the physical space, affecting physical system behavior.

Similar to the described closed-loop control mechanism, the *managing system* senses the current IPC resource and service states using probes and monitoring services. The DT services replicate these states into the virtual space. By this, also the physical system states can be replicated since they are already stored for processing within IPC resources. The so-created CPS virtual instances (i.e., the Dev DT and Ops DT) accurately reflect the entire CPS context (i.e., the CPS physical instance) to support CPS monitoring, control, maintenance, and adaptation. The software agents replicate virtually applied adaptations back into the physical space using effectors, thereby adjusting IPC and physical system states and behavior.

**DT-enabled Context.** The described DT-enabled close-loop mechanism can establish the context-awareness required by the managing and managed systems for autonomous CPS adaptations. As already discussed, the accuracy of this context depends on the twinning fidelity, which depends on the twinned states/parameters and the individual twinning rates of these parameters. In the CPS context, the twinning fidelity also depends on the requirements and capabilities of the AP levels, which is not reflected in the conceptual model shown in Figure 4. The DT context here focuses on accurately covering highly dynamic physical and virtual (i.e., IPC resource and service) states to accurately reflect the CPSs' underlying embedded mechatronic control infrastructure.

**DT-enabled DevOps.** The Dev and Ops *support processes* shown in Figure 4 interact with the Dev DT and the Ops DT to observe and adapt CPS behavior in virtual space. It, is the responsibility of the DT services to distribute and apply virtual changes to the physical CPS instance.

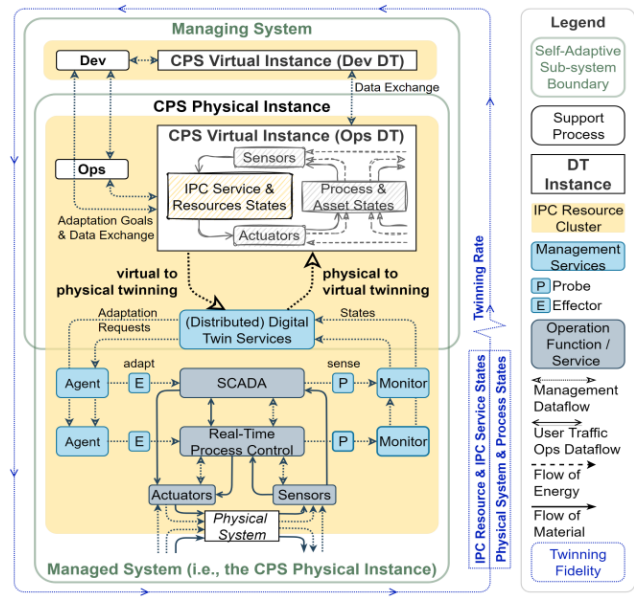

**Figure 4: Conceptual DT-enabled Self-Adaptive CPS Model**

## 4 Technical Self-Adaptive CPS Design Models

**Generic Device-Level Design Model.** Based on our previous work [53,56] and the analysis of the above-identified design requirements *R1-R4* against the AP, we conclude that the needed OT service and communication adaptations must be coordinated hierarchically across AP levels. However, the design requirements must be addressed on the device level to provide fine-granular CPS OT function and service adaptation.

The implementation of these properties on the device level heavily depends on the used operating systems (OSs) and supported virtualization technologies such as hypervisors, containers, and virtual machines. Since selecting these technologies depends on individual use case needs and hardware capabilities, we describe a generic device-level design model that can be tailored for specific implementation needs.

Figure 5 shows our proposed generic design model for OT devices. The model is structured into layers representing the OT stack. The top three layers implement the actual CPS functionality. The underlying layers provide the necessary IPC resources. OT and IT are strictly separated, represented by the usage of different OSs. A mediator controls and coordinates the interaction between OT and IT. The OT-side is responsible for CPS control and the twinning of operation parameters represented by the operations DT on the data layer. The IT-side is responsible for managing and configuring both IT and OT components. The management DT represents the current IT and OT IPC states to coordinate the adaptation on the device and between devices. Since this coordination is strictly separated from the time- and mission-critical OT functions, the use of IT tools and protocols is feasible. Management and operation only need to interact during system adaptation.

Following the decentralized adaptation patterns as of [57], effectors (E) and probes (P) need to be implemented on all layers for each service and function to enable context- and self-aware adaptation at runtime. The implementation of effectors and probes depends on the individual technologies and functions to be adapted. Hence, service and function design shall follow the Microservice architecture pattern for the Industrial Internet-of-Things (IIoT) [56] to facilitate independent service, effector, and probe development, as well as DevOps principles [8]. Next, we discuss the design of the individual layers from bottom to top.

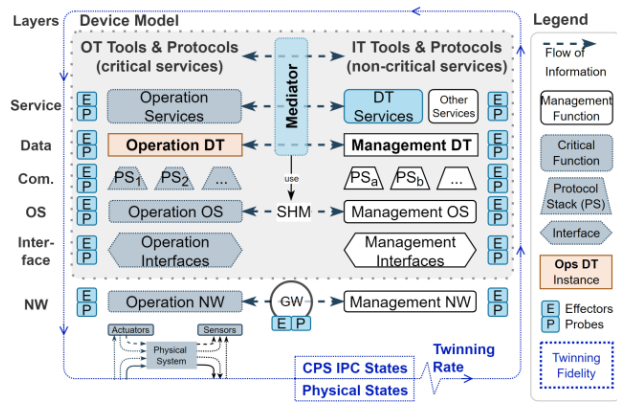

**Figure 5: Generic Self-Adaptive Device Design Model**

Like the device design, the network (NW) layer is separated into two communication planes, i.e., the operation and management NWs. Gateways (GWs) are commonly used as mediators and protocol translators between two incompatible NWs. These GWs also require effectors and probes for management as of [53,58]. The interface layer connects the device with the NW layers. The OS layer provides device hardware (HW) abstraction and management functions, and standard services for software development. The communication (com.) layer consists of one or more network and protocol stacks (PSs) to provide compatibility with multiple (fieldbus) network technologies. The data layer aggregates data based on the DT concept. The service layer implements the actual functionality using features of all lower layers. The mediator may span multiple layers coordinating the data flow and service access between operation and management.

Summing up the above design strategies, each device acts in principle like a software-controlled and software-defined data flow and service-provisioning gateway between the operation and management domain. Hence, design, management, and configuration strategies known from software-defined networking (SDN) [58,68] and the service mesh principle [53] are applicable. For example, the CoC pattern [58] is targeted toward the hierarchical management of heterogeneous networks, and the dependable mesh networking patterns [53] are targeted toward the integrated management of heterogeneous (Micro)services, networks, and devices. Both approaches propose using distributed controllers and hierarchical knowledge repositories for management.

**Runtime Adaptation & Verification Model.** Next, we apply the above strategies and models to transfer the A/B testing DS from the IT domain to the OT domain. In particular, our DT–enabled DevOps for CPS (as of Table 1) shall support: (i) Shadow deployment of new services for (ii) real-traffic testing (iii) on targeted users (i.e., specific assets and (sub-)processes) in the "real" CPS OT operation environment with the ability to (iv) operationalize the newly deployed services (after virtual and physical validation), and thereby (v) causing zero CPS downtime during the entire deployment procedure.

To that end, we integrate the generic device-level model into the architecture adaptation model. Figure 6 shows the resulting A/B testing model of distributed OT control services. In particular, the proposed model enables the verification of a (redesigned) control algorithm (i.e., the B service) by deploying B to the operation environment. At the same time, B's behavior and performance shall be compared to the currently active control algorithm (i.e., the A service). To that end, the IPC resources required by B must be allocated and configured, and B's network and inter-service communication links must be established while A continues its normal operation. B shall start its shadow operation aside of A after configuring B's IPC resources and services. The hierarchical DT services are responsible for managing the deployment (see Section 3).

CPS integrity shall be preserved by ensuring freedom of interference between A and B. Therefore, the signals from B to the actuators are blocked by software and not forwarded to the fieldbus network, ensuring that A is the only service controlling the physical system. Whereas, A and B receive the same sensor signals and can interact with and use all IPC resources above the fieldbus level. At these higher AP levels, freedom of interference is provided by separating not only the MDFs and ODFs but also the individual dedicated ODFs of A and B as shown in Figure 6 by using, e.g., priority-based flow control and time-triggered communication technologies [69].

The DT-based management services ensure integrity and availability by the autonomous coordination of the adaptation sequence. In addition, they continuously monitor the entire adaptation sequence and the A/B shadow operation (a) to abort eventually and roll back applied changes preserving CPS availability and integrity, and (2) to record and twin A, B, IPC, and CPS behavior and performance. Dev and Ops can use the twinned data for live and historical analysis in the design space.

CPS services in the OT domain typically operate in a cycling and time-synchronous manner, which allows the implementation of relatively simple time-triggered software-switching mechanisms to activate B and deactivate A within the same processing cycle. Therefore, A and B must agree on a specific point in time (i.e., a specific processing cycle) where that switch is set for A and B simultaneously, which enables switching services A and B during operation. Since A can continue its operation beside B, a rollback can be performed using the same switch mechanism.

**Implementation Model.** Next, we implement the above models on an OT control device. Therefore, we propose the pipeline-based IPC model shown in Figure 7. Figure 8 shows the corresponding task scheduling diagram for A/B testing.

The dataflow (DF) shown in Figure 7 is separated into a so-called management plane (MP) and operation plane (OP). Each plane is encompassed by the indicated twinning fidelity, i.e., the twinned parameters and the twinning rate. The tasks (i.e., circles representing threads, processes, and containers) on the OP within the blue box are operated in a cyclic time-synchronous mode, as shown in Figure 8. All other tasks outside that box operate in an asynchronous mode and can run whenever CPU resources are free. The protocol stacks, located on the subscriber and publisher sides, also represent tasks.

The input and output flow tasks are responsible for dataflow dispatching, monitoring, and priority-based flow control. The asynchronous management services (i.e., DT services as of Figure 5) can push adaptation and configuration requests to the operation plane's input flow control. These requests are forwarded to the cyclic input processing tasks. These tasks primarily receive data from the OP, besides the requests from the DT service from the MP.

As of Figure 8, the input processing tasks are the first tasks executed at each cycle start $t_0$, indicating the current point in time. Other devices and tasks on the OP operate synchronously to cycle $t_0$, which is achieved via hardware-supported distributed time-synchronization mechanisms. This synchronous operation is essential for accurate real-time closed-loop control and data exchange between OT services.

In principle, an execution cycle consists of four stages as of Figure 7: (1) (Pre-)processing and transfer of all input signals (i.e., subscriptions) to the data layer, where they are stored in the local high fidelity DTs. (2) Execution of the control algorithms that read the input signals and write the resulting control signals back to the DT in virtual space. (3) Forwarding of the closed-loop control signals to their specific destinations, such as actuators and other control devices. (4) Execution of asynchronous tasks like management, visualization, and logging tasks. These four principle steps bring us to the detailed discussion of our design and execution sequence. However, before going into the details, we make a short excurse to explain the underlying synchronization principle. As of Figure 5, the mediator is responsible for service and data layer synchronization. We decided to use the Disruptor pattern [67] as a basis for its implementation. The Disruptor is a bounded ring buffer data structure, as shown in the legend of Figure 7. It is designed to replace queues as synchronization mechanisms in multi-threaded producer-consumer applications that follow the event-based pipeline execution pattern. For such applications, the Disruptor enables high throughput ratios, low latency, and low jitter [70]. The Disruptor achieves this by using sequence counters as the only synchronization mechanism between threads, combined with a CPU-friendly memory layout and memory access pattern to exploit modern multi-core CPU hardware features. Further details on these mechanisms and analysis are given in [70] and [71].

In our above-described cyclic application, each cycle start denotes an event and multiple tasks are executed in sequence to operate on the same data entry (i.e., buffer entry), which perfectly matches the pipeline principle. However, the

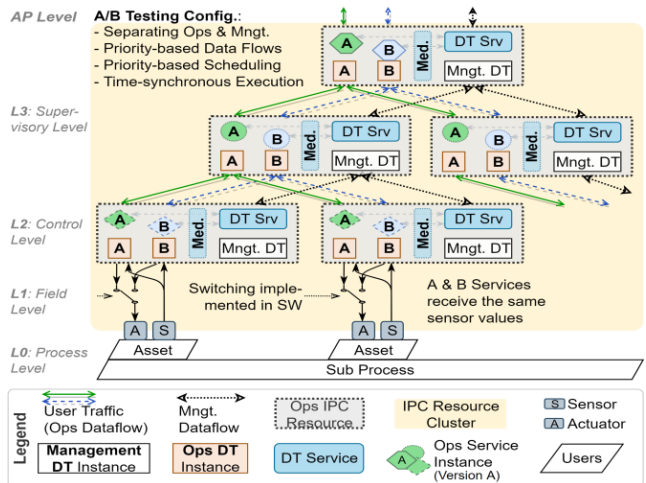

Figure 6: A/B Testing Model for OT Environments

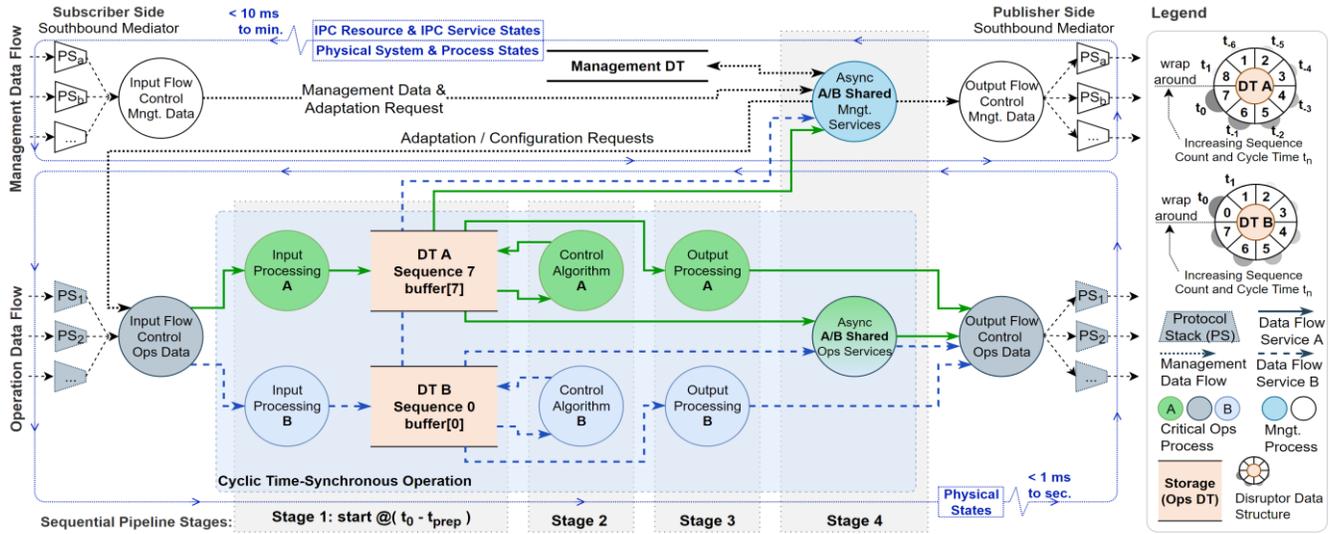

**Figure 7: Implementation Model (i.e., Dataflow Model) of a DT-enabled Architecture-based Self-Adaptive Control Device**

Disruptor structure needs to be adapted to meet our OT and adaptation requirements: (1) We must provide service and function adaptation at runtime, and (2) we must ensure that (producer) tasks on the OP, such as the input processor tasks, are not blocked if slower tasks (most likely on the MP) cannot take pace with the production and processing speed of the OP.

Addressing point (1), we implemented the Disruptor as a shared-memory data structure, which enables adding and removing pipeline stages and tasks before every cycle starts. To that end, the management services can send an adaption request. As of Figure 8, the input tasks process this request shortly before (i.e., **t$_{prep}$** for preparation time) the actual cycle start, which enables altering the individual pipeline stages and tasks in every cycle. *The so-enabled adaptation of tasks and their execution order allows seamlessly changing the provided control device functionality at runtime.*

Addressing point (2), we modified the original Disruptor synchronization mechanism (i.e., the barrier mechanisms as of [71]) that prevents producers from overtaking consumers. We replaced the initially blocking mechanisms with a wait-free non-blocking mechanism. This mechanism ensures that the input processing tasks can always reserve a buffer entry for the next cycle. To that end, we introduce a transaction-based synchronization mechanism for consumers, under the premise that all consumers after pipeline stage 2 are only allowed to read from the Disruptor memory buffer. Whenever a producer (i.e., the input processing tasks) reserves a buffer entry, this entry is invalidated via a wait-free update (i.e., store-release instruction) of the cycle timestamp. This ensures that the producer is not blocked from making progress while all (asynchronous) consumers recognize the entry invalidation.

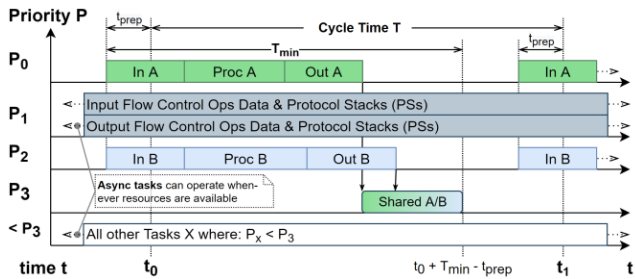

**Figure 8: Scheduling Diagram A/B Testing**

In the previous sections, we identified availability and integrity as essential for CPS. However, every adaptation at runtime introduces uncertainty and may result in availability and integrity violations. To minimize device-level adaptation risks, we first assign newly deployed operation tasks to a lower scheduling priority class. Hence, service B tasks are assigned to the class $P_2$, as of Figure 8, which shall ensure on a modern multi-core CPU (in our analysis, we assumed four cores) that all other tasks can make progress, even in the case that B blocks an entire CPU core. Service B is terminated if management services detect IPC resource usage limit violations. Otherwise, the operator can decide to trigger the above described switching mechanisms to seamlessly deactivate A while making B the active service to interact with the physical system level.

## 5 Evaluation and Discussion

Regarding coverage of research questions *RQ1* to *RQ4*, we can report as follows: As for RQ1, we derived design requirements of self-adaptive CPS IPC services and their DT(s), starting from a value-driven perspective of IPSS service needs on their underlying CPS IPC platform. This led us to the key requirements related to context-aware self-adaptability while assuring CPS availability and integrity.

Integrating context-awareness is achieved via the layered and distributed DT design that links all AP levels. The so established DT context is not limited to static and slowly changing physical properties and states like in traditional DT concepts. Instead, our DT context focuses on covering highly dynamic physical and virtual states to accurately reflect the CPSs' underlying embedded mechatronic control infrastructure, including its IPC resources and services. Such high fidelity is achieved by distributing and integrating the modular DT instances at all CPS scales (i.e., system, device, service) and AP levels. By this, every DT instance can operate at its individual fidelity for its dedicated purpose. Multiple DT instances agree on a common fidelity for data exchange by that establishing an adaptive CPS context across AP levels. As for *RQ2*, such a design enables a high-fidelity adaptive context across AP levels.

As for *RQ3*, we have developed a conceptual model of a DT-enabled self-adaptive CPS that allows replicating the entire physical CPS context (i.e., physical asset, IPC resource, and IPC service states) into the virtual space and back again. This bidirectional twinning of the CPS context can be used by Dev and Ops teams for CPS monitoring, analysis, and adaptation.

As for *RQ4*, we instantiated the above conceptual model and demonstrated the feasibility of the proposed context- and self-aware deployment, adaptation, and A/B testing on a evaluation platform for distributed CPS control. From our perspective, the most relevant part of the demonstration is that our proposed DT concept can serve as a distributed knowledgebase for hierarchical coordinated CPS adaptation. In particular, we showed that complex CPS functions that span multiple AP levels could be deployed to the Ops space for validation without causing CPS interferences and downtime. From a design view, our results reveal that the proposed DT-enabled self-adaptive CPS models can be instantiated for specific real-world use cases, i.e., the A/B testing of services at device and system scale while meeting the strict OT availability, integrity, and real-time constraints of sub-millisecond cycle times. The demonstrated DS also shows that a modular design enables the DevOps-like closed-loop optimization of CPS service changes through architecture-based self-adaptation and verification at runtime.

The proposed models come also with challenges in terms of implementation complexity. Compared to traditional solutions, they require more IPC resources in terms of processing power, memory usage, and communication bandwidth to implement their functionality. Hence, future extended evaluation activities include investigating detailed timing, maximum tolerable IPC resource usage on device, NW, and system scale, and a detailed adaptation behavior analysis under uncertain CPS conditions.

## 6  Conclusion

Guided by the CPS IPSS Automation Pyramid architecture and the IT DevOps concept, this article proposes generic technical design models that facilitate the implementation of adaptive and evolvable IPSS based on DT's. The use case that served as a design objective is A/B testing of CPS IPSS service changes in the operation-critical CPS environment at runtime. The main contribution of this work is to ensure full CPS IPSS availability and integrity during CPS adaptation and A/B testing. This contribution is pivotal for transferring the DevOps concept known in the IT domain to the OT domain, which opens the door for entirely new opportunities in terms of bringing closely together IPSS design and adaptive operation, both from a technical and organizational perspective.

## ACKNOWLEDGMENTS

The authors thank Andritz Hydro GmbH and the Austrian Research Funding Agency FFG for supporting this research.